\documentclass[]{aa}
\usepackage{hyperref}
\usepackage{graphicx}
\usepackage{txfonts}
\usepackage{color}

\begin{document}

   \title{Horizontally and vertically polarized kink oscillations in curved solar coronal loops}

  \titlerunning{Kink oscillations in curved coronal loops} 
  
   \author{Mingzhe Guo\inst{1,2}
          \and
          Tom Van Doorsselaere\inst{1}
          \and
          Bo Li\inst{2}
          \and
          Marcel Goossens\inst{1} 
          }

   \institute{Centre for mathematical Plasma-Astrophysics (CmPA), 
   Department of Mathematics, KU Leuven, Celestijnenlaan 200B, 
   B-3001 Leuven, Belgium\\
   \email{Mingzhe.Guo@kuleuven.be}
   \and
    Shandong Provincial Key Laboratory of Optical Astronomy and 
    Solar-Terrestrial Environment,
    Institute of Space Sciences, Shandong University, Weihai 264209, China
               }
   \date{Received; accepted}

 
  \abstract
   {}
   {Kink oscillations are frequently observed in coronal loops. 
   This work aims to numerically clarify the influence of loop curvature on horizontally and vertically polarized kink oscillations.}
   {Working within the framework of ideal magnetohydrodynamics (MHD), we conduct three-dimensional (3D) simulations of axial fundamental kink oscillations in curved density-enhanced loops embedded in a potential magnetic field.
   Both horizontal and vertical polarizations are examined, and their oscillation frequencies are 
        compared with WKB expectations.
   We discriminate two different density specifications. 
   In the first (dubbed ``uniform-density''), the density is axially uniform and varies continuously in the transverse direction toward a uniform ambient corona.
   Some further stratification is implemented in the second specification (dubbed ``stratified''), 
       allowing us to address the effect of evanescent barriers.
   }
   {Examining the oscillating profiles of the initially perturbed uniform-density loops,
   we found that the frequencies for both polarizations deviate from the WKB expectation by $\sim 10\%$.
   In the stratified loop, however,
   the frequency of the horizontal polarization deviates to a larger extent ($\sim 25\%$).
   We illustrate the lateral leakage of kink modes through wave tunnelling in 3D simulations, for the first time.
   Despite this, 
   in both uniform-density and stratified loops, 
   the damping time-to-period ratios are similar and close to the analytical predictions for straight configurations under the thin-tube-thin-boundary (TTTB) assumption.
   }
   {The WKB expectation for straight configurations can reasonably describe the eigenfrequency of kink oscillations only in loops without an asymmetrical cross-loop density profile perpendicular to the oscillating direction.
   Lateral leakage via wave tunnelling is found to be less efficient than resonant absorption,
       meaning that the latter remains a robust damping mechanism for kink motions even when loop curvature is included.}

   \maketitle

\nolinenumbers

\section{Introduction}

In recent years,
more and more observations have revealed that 
magnetic structures in the solar corona are dynamic and support
a rich variety of low-frequency waves in the magnetohydrodynamic (MHD) regime \citep[see e.g.,][for recent reviews]{2020SSRv..216..136L,2021SSRv..217...34W,2021SSRv..217...73N,2021SSRv..217...76B}.
As energy carriers, these waves may be of potential importance for tackling the coronal heating problem \citep[see the reviews by e.g.,][]{2015RSPTA.37340269D,2015RSPTA.37340261A,2020SSRv..216..140V}.
Likewise, these waves themselves carry information about their hosts,
and thus have been considered useful for probing the physical conditions in the 
solar corona \citep[e.g.,][]{2005LRSP....2....3N,2012RSPTA.370.3193D,2020ARA&A..58..441N}.
Furthermore, coronal waves are believed to be related to the energy release
processes in solar flares as well \citep[e.g.,][]{2018SSRv..214...45M,2021SSRv..217...66Z,2023BAAS...55c.181I}.

Canonically accepted as the only motions that displace structural axes, 
kink modes have been frequently imaged in the solar corona
(see e.g., \citealt{2021SSRv..217...73N} for a recent review; 
see also \citealt{1999Sci...285..862N,1999ApJ...520..880A} for early dectections).
They have been routinely employed in coronal seismology, 
which combine wave theories and observations to
deduce those coronal parameters that are difficult to directly measure.
For instance,
the measured periods of kink motions have been in routine use for inferring the coronal magnetic field
strength \citep[e.g.,][]{2001A&A...372L..53N,2020Sci...369..694Y}.
Likewise, transverse inhomogeneity lengthscales can be seismologically deduced
    with the measured damping times of decaying kink motions, 
    provided that this damping is attributable to such mechanisms as 
    resonant absorption \citep[e.g.,][]{2002A&A...394L..39G,2003ApJ...598.1375A,2014A&A...565A..78A,2019A&A...625A..35A}.
On this aspect, we note that resonant absorption is an ideal process that
    transfers kink energy into localized Alfv\'enic motions
    (see the review by \citealt{2011SSRv..158..289G}).
This process has been shown to be robust for a rich set of configurations  
    \citep[e.g.,][]{2004A&A...424.1065V,2008ApJ...679.1611T,2011ApJ...731...73P,2017A&A...607A..77H,2020ApJ...904..116G, 2024arXiv240211181S}.

Solar coronal loops have been customarily modelled as straight magnetic cylinders
    in theoretical studies on kink motions
    (e.g., \citealt{1983SoPh...88..179E}; see also \citealt{1975IGAFS..37....3Z,1979A&A....76...20W,1982SoPh...75....3S,1986SoPh..103..277C}). 
For instance, resonantly damped kink motions have been extensively
    studied from the ideal quasi-mode perspective with 
    the Frobenius approach \citep[e.g.,][]{2013ApJ...777..158S,2019A&A...623A..32S,2022A&A...661A.100G}
    or more frequently dissipative eigenmode computations
    \citep[e.g.,][]{1991PhRvL..66.2871P,2004ApJ...606.1223V,2006ApJ...642..533T,2016SoPh..291..877G}.
In particular, concise expressions are available for the oscillation frequencies
    and damping rates of kink quasi-modes when the thin-tube-thin-boundary (TTTB) approximation applies \citep[e.g.,][]{1992SoPh..138..233G,2009A&A...503..213G,2013ApJ...777..158S}.
From the initial value problem perspective,
    it has been shown that the theoretical expectations for kink quasi-modes
    tend to well describe the temporal evolution of kink motions
    \citep[e.g.,][]{2002ApJ...577..475R}, 
    particularly when loop boundaries are thin \citep{2015ApJ...803...43S}.

Closed loops, believed to be typical magnetic structures in the corona \citep[][]{2014LRSP...11....4R},
serves as the main waveguide for kink motions.
A question then arises how does the loop curvature influence the eigenfunctions of kink modes?
Previous wave-related investigations focused on two-dimensional 
(2D) loop models \citep[e.g.,][]{1997A&A...317..752S,2005A&A...438..733B,2005A&A...440..385S,2006A&A...454..653S,2006A&A...446.1139V,2006A&A...449..769V,2006A&A...452..615V}.
In particular,
\cite{2006A&A...446.1139V,2006A&A...449..769V,2006A&A...452..615V}
analyzed fast magnetoacoustic modes and their seismological potential in detail in a curved slab model under the plasma $\beta=0$ limit.
Lateral leakage of kink motions via wave tunnelling has been examined.
However,
these two-dimensional studies are constrained to
vertically polarized kink modes only.
While vertical polarizations have been reported \citep[e.g.,][]{2004A&A...421L..33W},
a more comprehensive examination,
including the more frequently observed 
horizontal polarization \citep[e.g.,][]{2023NatCo..14.5298Z},
should also be taken into account.
This can only be achieved in three-dimensional (3D) examinations.
Initial analytical progress has been made by \citet{2004A&A...424.1065V,2009SSRv..149..299V}
in a toroidal model embedded in a radially dependent force-free field,
showing that the curvature of a loop has little influence 
on the kink eigenmodes in the absence of wave leakage.
Beyond the consideration of a constant loop cross-section,
\citet{2009A&A...506..885R} found that the difference in eigenfrequencies
of vertical and horizontal oscillations is proportional to the 
tube expansion parameter under the thin tube (TB) approximation.
Numerically,
the semi-torus model with a constant loop cross-section considered by \citet{2006ApJ...650L..91T} distinguished 
horizontally and vertically polarized oscillations.
The frequency difference between these two polarizations is not observable.
Meanwhile,
lateral leakage that is naturally induced in curved loops 
has also been included,
and it is found to be less efficient than resonant absorption.
The response of such a semi-circular loop model to external perturbations was also numerically examined by \citet{2014ApJ...784..101P}.
In addition,
kink oscillations in a force-free magnetic field with a dipole configuration have been considered by e.g., \citet{2008ApJ...682.1338M,2020ApJ...894L..23M}.
In particular,
strong damping of kink oscillations is illustrated by \citet{2008ApJ...682.1338M},
and it was attributed to the curvature of their loop
model.
It seems that the efficiency of lateral leakage and resonant damping
in curved loops still needs to be further clarified.
Regarding the frequency of kink modes in curved loops,
\citet{2020ApJ...894L..23M} found that the WKB approximation can reasonably describe the kink period in their planar loop model without considering a sigmoid geometry.
However,
the influence of a potential field configuration on the different polarizations of kink modes still remains unclear.

In the current work,
we consider a curved loop model embedded in a potential magnetic field and examine its response to 
initial velocity perturbations from different directions.
The influence of loop curvature on the eigenfrequency and damping
rate of the excited kink polarizations will be examined.
Our study differs from available studies in the following aspects. 
Firstly,
both horizontal and vertical polarizations will be considered and their frequencies will be compared with the WKB approximation.
Secondly,
lateral leakage will be investigated in 3D simulations,
and the efficiency of this damping mechanism will be examined and compared with resonant absorption. 
This paper is organized as follows. 
Section \ref{sec_model} describes the equilibrium configuration and the numerical setup. 
The simulation results are presented in Section \ref{sec_results}, 
followed by a summary and discussion of the present study in Section \ref{sec_summary}.

   \begin{figure*}
   \centering
   \includegraphics[width=1.\hsize]{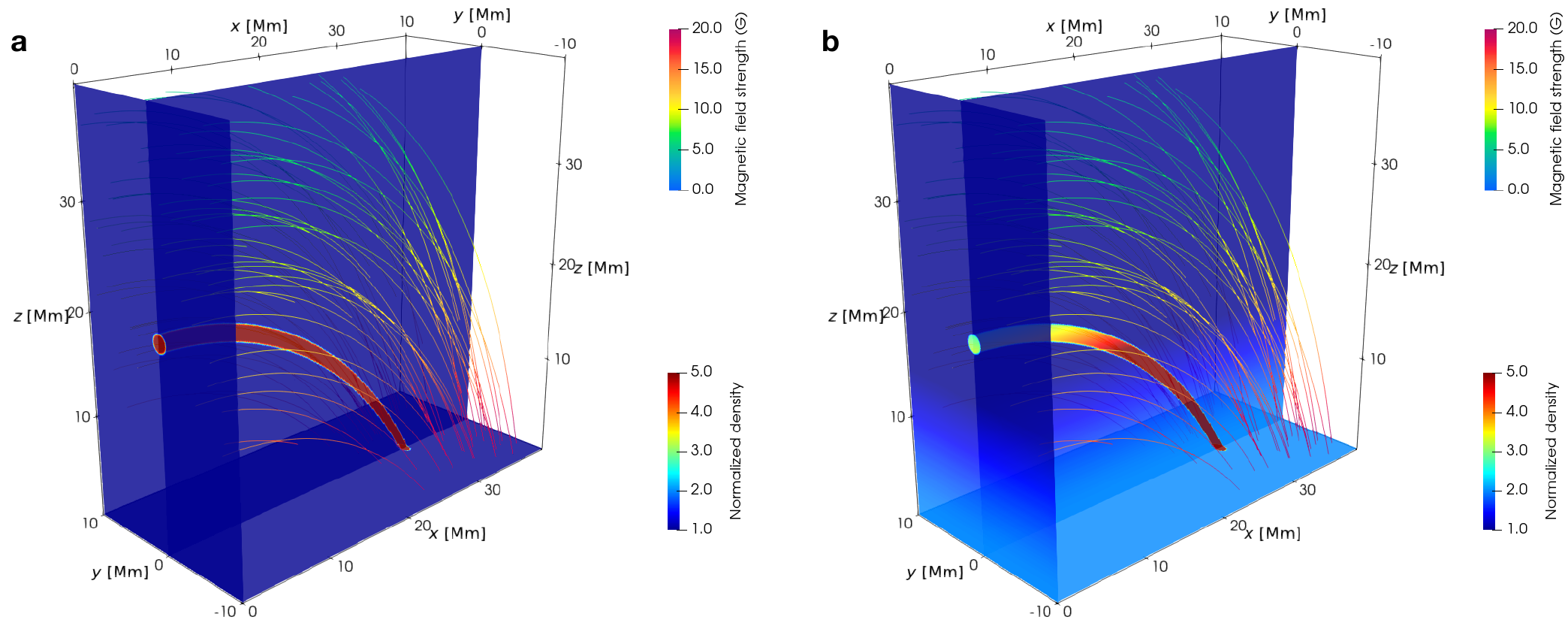}
   \caption{Three-dimensional rendering of the density-enhanced loop.
   Several arbitrarily selected magnetic field lines are also shown and color-coded by the local magnetic field strength. The left and right panels correspond to our ``uniform-density''
   and ``stratified'' computations. See text for more details. 
   }
    \label{fig_loop_snapshot}
    \end{figure*}

\section{Numerical Model}
\label{sec_model}

We consider a potential magnetic field to mimic
the magnetic structures in the solar corona.
A similar configuration has been considered by
e.g., \citet[][]{1993A&A...273..647O,2005A&A...440..385S,2013ApJ...763...16R}.
Let $(x,y,z)$ be a Cartesian coordinate system.
The equilibrium magnetic field is given by 
\begin{eqnarray}
B_x(x,z) = B_0\cos\left(\displaystyle\frac{x}{\Lambda_B} \right)\exp\left(-\displaystyle\frac{z}{\Lambda_B} \right),\\
B_z(x,z) = -B_0\sin\left(\displaystyle\frac{x}{\Lambda_B} \right)\exp\left(-\displaystyle\frac{z}{\Lambda_B} \right),
\label{eq_B}
\end{eqnarray}
where $B_0=20$G,
$\Lambda_B=2L_B/\pi$ with $L_B=45$Mm.
This magnetic field can be alternatively expressed via the vector potential
\begin{eqnarray}
\vec{A} = \psi \hat{y}=B_0\Lambda_B\cos\left(\displaystyle\frac{x}{\Lambda_B}\right)\exp\left(-\displaystyle\frac{z}{\Lambda_B}\right)\hat{y},
\label{eq_psi}
\end{eqnarray}
    where $\psi$ is the magnetic flux function whose contours delineate magnetic lines of force.
A set of right-handed orthonormal basis vectors $(\hat{\vec{e}}_t, \hat{\vec{e}}_h, \hat{\vec{e}}_v)$
   can be defined, where
\begin{equation}
\label{eq_basisVec}
\begin{split}
& \hat{\vec{e}}_{\rm t} = \cos(x/\Lambda_B)\hat{x}-\sin(x/\Lambda_B)\hat{z},  \\ 
& \hat{\vec{e}}_{\rm h} = \hat{y},                                            \\
& \hat{\vec{e}}_{\rm v} = \sin(x/\Lambda_B)\hat{x}+\cos(x/\Lambda_B)\hat{z}. 
\end{split}
\end{equation}
Evidently, $\hat{\vec{e}}_{\rm t}$ ($\hat{\vec{e}}_{\rm v}$) is locally 
    parallel (perpendicular) to the magnetic field.
As such, we further identify $\hat{\vec{e}}_{\rm h}$ and $\hat{\vec{e}}_{\rm v}$ 
    as characterizing the horizontal and vertical polarizations, respectively.

Density-enhanced loops are constructed as follows.
Consider, for now, a uniform-density loop, by which we mean the density 
   is uniform along the magnetic field.
The density distribution is given by
\begin{eqnarray}
\rho(\Bar{r}) = \rho_
{\rm e}+(\rho_{\rm i}-\rho_{\rm e})f(\Bar{r}),
\label{eq_rho}
\end{eqnarray}
with
\begin{eqnarray}
f(\Bar{r}) = \displaystyle\frac{1}{2}\left\{1-\tanh\left[\left(\displaystyle\frac{\Bar{r}}{R}-1\right)b\right]\right\},
\label{eq_f}
\end{eqnarray}
where
\begin{eqnarray}
\Bar{r}= \sqrt{\left(\displaystyle\frac{\psi-\psi_0}{\Delta\psi}\right)^2+\left(\displaystyle\frac{y}{\Delta y}\right)^2},
\label{eq_r}
\end{eqnarray}
with $\psi_0=\psi(x=30{\rm Mm},z=0)$,
$\Delta\psi=B_0\Lambda_B$,
$\Delta y=2\Lambda_B$.
Here $\rho_{\rm i}$ ($\rho_{\rm e}$) represents the internal (external) density,
and $R={\cos\left(30/\Lambda_B\right)-\cos\left(30.5/\Lambda_B\right)}$ prescribes
   the $x$-extent of the loop at its footpoints. 
The parameter $b$ determines the width of the boundary layer.
The 3D view of the loop can be seen in
Figure~\ref{fig_loop_snapshot}a.

We proceed to specify the rest of the physical parameters.
The internal (external) density is $\rho_{\rm i}=1.17\times10^{-14}{\rm g}~{\rm cm}^{-3}$
($\rho_{\rm e}=2.34\times10^{-15}{\rm g}~{\rm cm}^{-3}$).
The thickness of the boundary layer is determined by $b=10$,
which gives the layer width $l\approx 0.3a$ with $a$ being the half-width of the cross-section at the loop apex.
The temperature of the background corona is $T_{\rm e}=1$MK.
The pressure is uniform in the entire computational domain,
which is set to be $p=2.3k_{\rm B}\rho_{\rm e} T_{\rm e}$,
with $k_{\rm B}$ being the Boltzmann constant.

To distinguish between different polarizations of kink modes,
we employ two initial velocity perturbations.
To excite the horizontal polarization,
we introduce an initial velocity perturbation of the form
\begin{eqnarray}
  \vec{v}(x,y,z; t=0) 
= v_0\sin{\theta}f(\Bar{r}) \hat{\vec{e}}_h,
\label{eq_vh}
\end{eqnarray}
where $v_0$ represents the amplitude of the velocity perturbation,
$\sin\theta=\dfrac{z}{\sqrt{x^2+z^2}}$ with $\theta$ being the angle along the loop from the footpoint,
$f(\Bar{r})$ is given by Equation \eqref{eq_f}.
It ensures maximum velocity perturbation at loop apex and zero velocity at loop footpoints in the $y$-direction. 
Similarly, 
we introduce the other velocity perturbation to excite the vertical polarization, which reads
\begin{eqnarray}
  \vec{v}(x,y,z; t=0) 
= v_0\sin{\theta}f(\Bar{r}) \hat{\vec{e}}_v.
  \label{eq_vv}
\end{eqnarray}
To limit the simulations to the linear regime,
we set the velocity amplitude to 
 ${v_0=5 {\rm km/s}}$.
 This choice ensures that the maximum displacement of the loop is much less than the loop radius. 
The velocity fields of the initial perturbations are shown in Figure~\ref{fig_vh}a and Figure~\ref{fig_vv}a.
Although these initial perturbations can not exactly match the form of the eigenmodes,
which are not straightforward to obtain beforehand in the current model,
kink eigenmodes are subsequently excited in our loop models as a response to these initial velocity perturbations. 

The boundary conditions are specified as follows.
Reflective boundary conditions are applied for 
   the three components of velocity at $z=0$ to mimic the footpoints of the magnetic structures anchored in the lower solar atmosphere.
All other variables at $z=0$ are set to have zero-gradients.   
Asymmetric boundary conditions are used for $v_x$, $B_y$, and $B_z$ to follow the oscillating properties of kink modes at $x=0$.
All other variables there are set to be continuous.
Zero-gradient boundary conditions are employed 
at the other side boundaries and the top boundary.

We evolve the set of 3D, ideal MHD equations with 
   the finite-volume code MPI-AMRVAC \citep{2018ApJS..234...30X, 2020arXiv200403275K}.
Gravity is neglected throughout.
As described in \cite{2017A&A...603A..42X},
we split the magnetic field into $\vec{B}=\vec{B_0}+\vec{B_1}$,
    taking $\vec{B_0}$ as the equilibrium one given by Equation~\eqref{eq_B}
    and actually solving for the time-dependent component ($\vec{B_1}$).
An approximate HLL Riemann solver and a third-order ``cada3" 
   slope limiter \citep{2009JCoPh.228.4118C} are employed 
   for evaluating inter-cell fluxes.
We use the third-order TVD Runge-Kutta method
   for time marching with a Courant number of $0.5$.
The computational domain is chosen to be $[0,40]\times[-10,10]\times[0,40]$Mm,
   which incorporates only the $x\ge 0$ portion given the symmetric properties
   of our numerical implementation. 
Adaptive mesh refinement (AMR) is employed to further reduce the computational cost.
We adopt a base grid of $40\times60\times120$.
Four levels of AMR are implemented, leading to the highest resolution
    of $41.7$km across the loop region.


\section{Results}
\label{sec_results}
\subsection{General snapshots}
\label{subsec_snapshots}

   \begin{figure*}
   \centering
   \includegraphics[width=0.8\hsize]{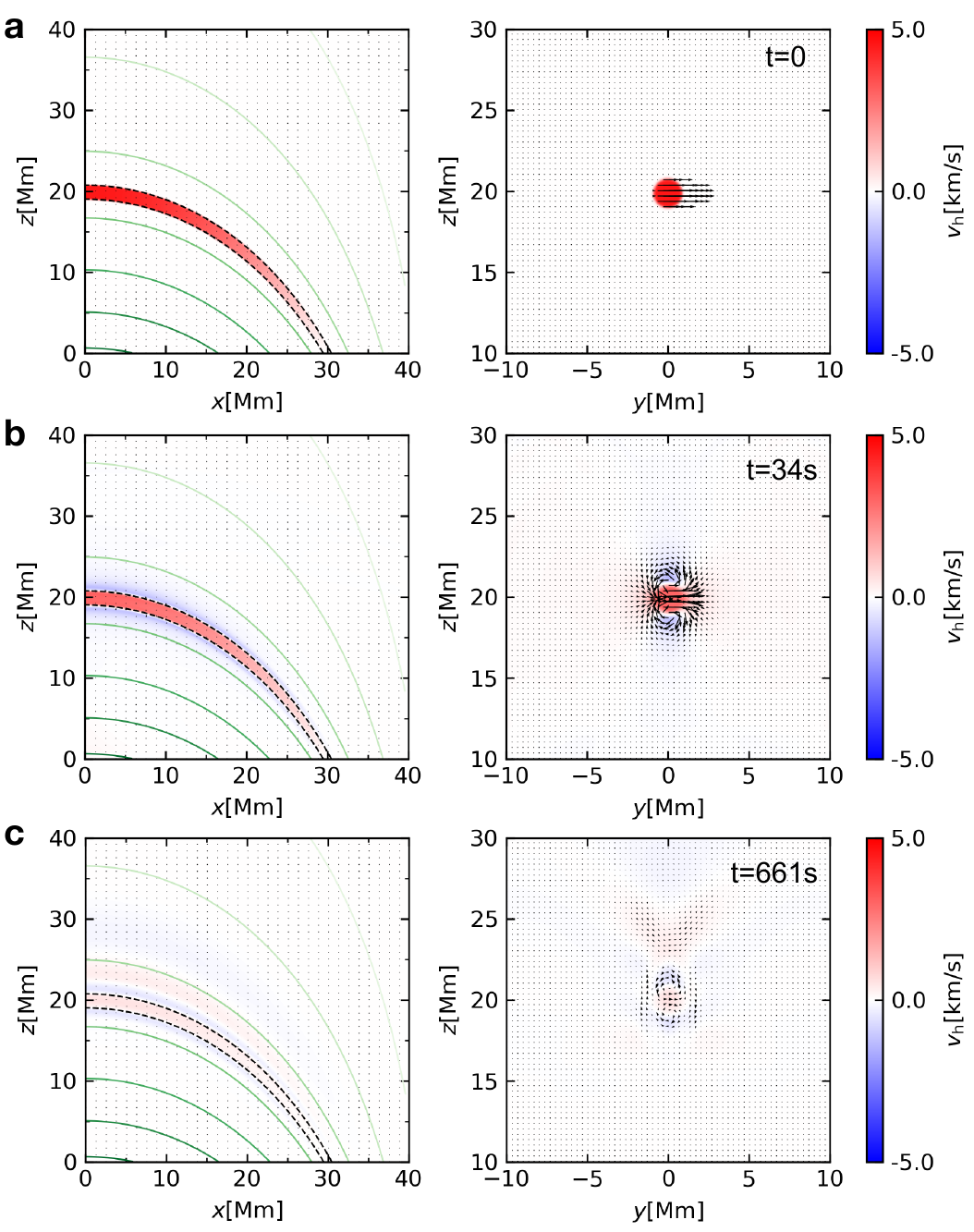}
   \caption{
   Velocity fields (arrows) for the horizontal polarization at
       (a) $t=0$, (b) $t=34$s, and (c) $t=661$s. 
   The left and right columns correspond to the $y=0$ and $x=0$ cuts, respectively.
   Overplotted are the filled contours of $v_{\rm h} = v_y$.
   The green curves in the left column represent the magnetic lines of force, 
       with the dashed lines further outlining the density-enhanced loop. 
   These snapshots are taken from the attached animations, 
       which run from $t=0$ to 1270s.}
    \label{fig_vh}
    \end{figure*}  

   \begin{figure*}
   \centering
   \includegraphics[width=0.8\hsize]{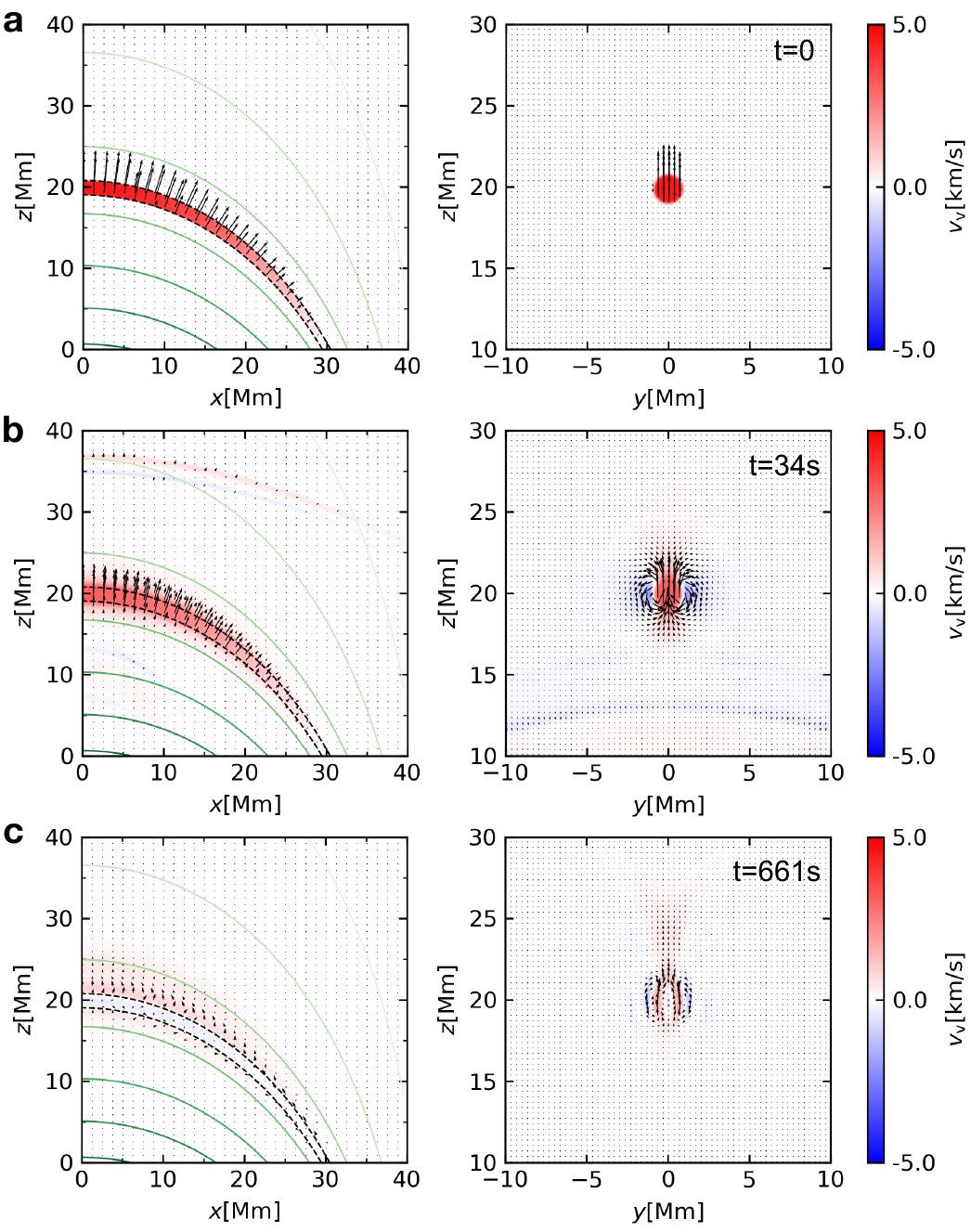}
   \caption{
   Similar to Figure~\ref{fig_vh} except that the filled contours are
      for $v_v\coloneqq \vec{v}\cdot\hat{\vec{e}}_v$.
   An animation is also attached, and runs from $t=0$ to 1270s.}
    \label{fig_vv}
    \end{figure*}

We first examine the response of the uniform-density loop 
   to the initial perturbations by analyzing the distribution of velocity in the loop.
Figure~\ref{fig_vh} and Figure~\ref{fig_vv} 
   show the velocity field at
   $y=0$ and $x=0$ (the loop apex) for horizontal and vertical polarizations at different times.
   The density-enhanced loop is outlined by dashed lines,
   and the magnetic lines of force are labelled by green lines.
We start by noting that wave reflection can be seen in Figure~\ref{fig_vv}b 
   due to the asymmetric boundary at the bottom surface.
However, the reflected fast waves quickly leave the computational domain
   and do not influence the following analysis.
From both Figure~\ref{fig_vh}b and Figure~\ref{fig_vv}b,
    we can clearly observe the well-known dipole-like velocity fields that are characteristic of kink motions \citep[e.g.,][]{2014ApJ...788....9G, 2020ApJ...904..116G} around the loop boundary.
From the related animations,
    we can observe damping kink oscillations in the loop region for both polarizations.
A signature of resonant absorption is identified as an increase
    of localized Alfv\'enic motions around the loop boundary.
For the horizontal polarization, Alfv\'enic motions can be found
    around $z=20.5$Mm and $z=19.5$Mm at the loop apex, namely the upper and lower boundaries of the loop cross-section.
Likewise, such Alfv\'enic motions can be observed around $y=-0.5$Mm 
    and $y=0.5$Mm at the loop apex for the vertically polarized mode.
Resonant absorption is thus one of the damping mechanisms in the current model.
This mechanism has been investigated in many previous works \citep[e.g.,][]{2020ApJ...904..116G}.
Here we demonstrate that resonant absorption is a robust damping mechanism
    that manifests itself in curved loops as well.
In addition to resonant absorption, lateral leakage of the wave modes 
    can also be observed in the animations and 
    Figure~\ref{fig_vh}(c) and Figure~\ref{fig_vv}(c) as selected snapshots.
Oscillations due to the wave leakage can be seen from the upper region outside
    the loop for both polarizations.
A detailed discussion of damping mechanisms will be presented shortly.


\subsection{Oscillating frequency}
\label{subsec_frequency}

   \begin{figure}
   \centering
   \includegraphics[width=1.0\hsize]{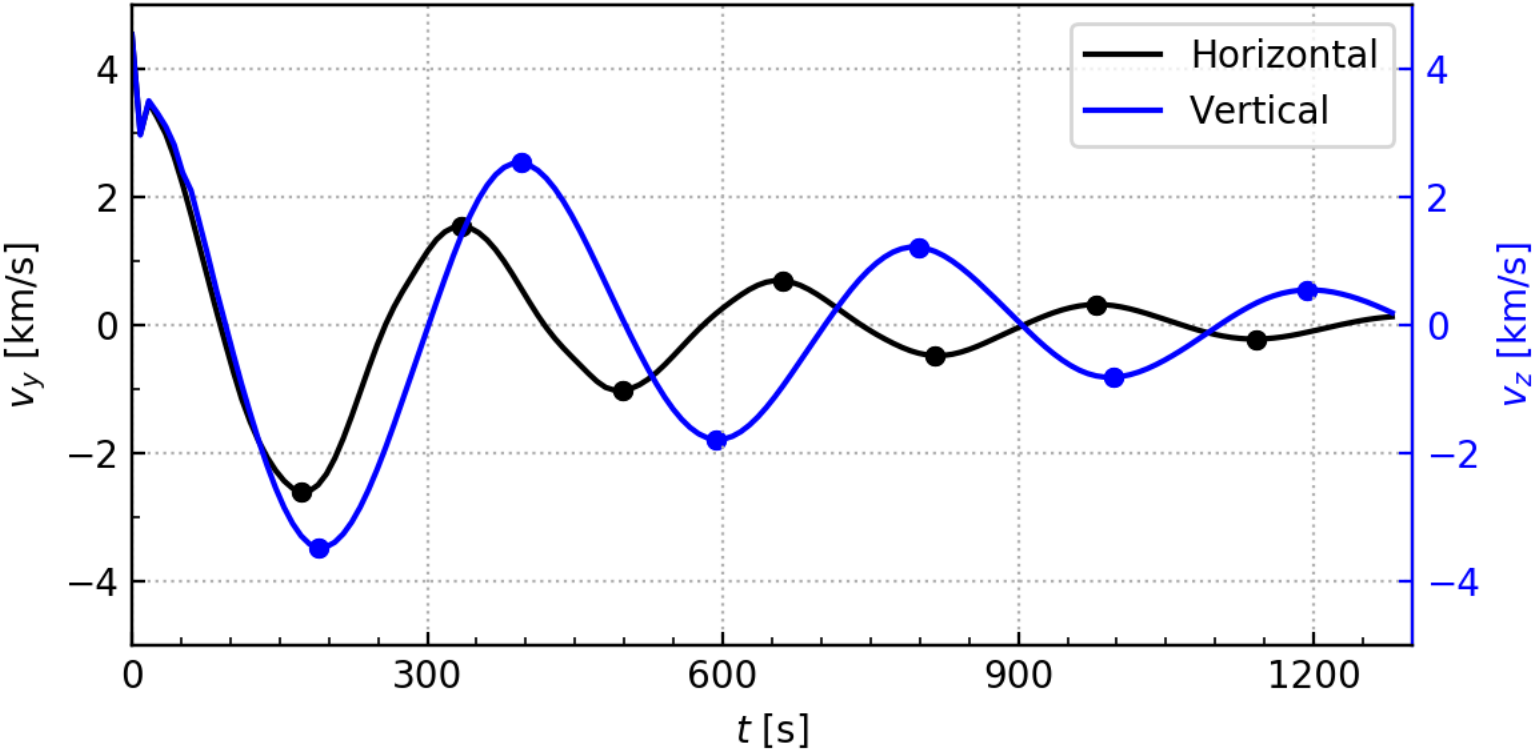}
   \caption{The oscillation profiles of the horizontally (black) and vertically (blue) polarized modes. The velocity is sampled at $x=0, y=0, z=19.85{\rm Mm}$. Solid circles represent the extrema of each curve.}
    \label{fig_vyz_t}
    \end{figure}  
    
We examine the oscillation frequency in both polarizations.
Figure~\ref{fig_vyz_t} shows the evolution of velocity sampled at 
    the loop apex for both polarizations.
We locate the local extrema in both oscillation profiles, 
    thereby evaluating the oscillation period as twice the average temporal spacing
    between two adjacent extrema.
Consequently,
we obtain two different periods for the horizontal polarization 
($P_{\rm h}=326.3$s) and the vertical case ($P_{\rm v}=401.8$s).

We can compare the afore-derived periods with analytical predictions.
For straight cylinders that are structured only in the radial direction,
    a simple expression is available for 
    the oscillation frequency of kink modes 
    in the TTTB limit \citep[e.g.,][]{1992SoPh..138..233G,2002A&A...394L..39G},
    writing
\begin{eqnarray}
\omega_{\rm k} = 
\sqrt{\displaystyle\frac{\rho_{\rm i} \omega^2_{\rm Ai}+\rho_{\rm e} \omega^2_{\rm Ae}}{\rho_{\rm i}+\rho_{\rm e}}},
\label{eq_TTTB_freq}
\end{eqnarray}
where $\omega_{\rm Ai}$ ($\omega_{\rm Ae}$) represents the internal (external) Alfv\'en frequency.
Equation~\eqref{eq_TTTB_freq} becomes even simpler when thermal pressure is negligible, 
    reading 
\begin{eqnarray}
\omega_{\rm k} = 
\omega_{\rm Ai}\sqrt{\displaystyle\frac{2\rho_{\rm ie}}{\rho_{\rm ie}+1}},
\label{eq_TTTB_freq2}
\end{eqnarray}
    where $\rho_{\rm ie}$ is the ratio of the internal to the external density.
Equation~\eqref{eq_TTTB_freq2} serves as a reasonable starting point for our further analysis,
    given the small value of the plasma $\beta$ in our equilibrium setup 
    ($\beta \lesssim 0.07$ in the loop region).
The internal Alfv\'en frequency changes with height,
we thus consider a WKB approximation to estimate
the eigenfrequency
\begin{eqnarray}
P_{\rm k}=2\sqrt{\frac{\rho_{\rm ie}+1}{\rho_{\rm ie}}}\int^L_0 \displaystyle\frac{ds}{v_{\rm Ai}(s)},
\label{eq_WKB}
\end{eqnarray}
with 
\begin{eqnarray}
v_{\rm Ai}(s)=\displaystyle\frac{B(s)}{\sqrt{\mu_0 \rho_{\rm i}}},
\label{eq_WKB_alfven}
\end{eqnarray}
where $L$ represents the loop length,
$s$ is the distance from a footpoint along the loop axis,
and $B(s)$ is the magnetic field strength along $s$.
Regardless of the damping first,
we can roughly estimate the eigenperiod of the current loop by considering Equation~\eqref{eq_WKB}.
The period can thus be readily obtained as $P_{\rm k}=369$s.
One can find that the horizontal period $P_{\rm h}$ has a deviation
of about $11.6\%$, compared with $P_{\rm k}$,
while the vertically polarized mode has a deviation
of about $8.9\%$.
It should be noted that these deviations are acceptable,
given the assumptions considered in deriving Equation~\eqref{eq_TTTB_freq2}.
Detailed discussions regarding the period discrepancy can be found in Section~\ref{subsec_freq_diff}.
Here we show that the frequency given by WKB theory is a good approximation to estimate the eigenfrequency of different kink polarizations in a curved loop.
Such a small frequency deviation from the WKB approximation in the curved loop model can be regarded as a positive result for coronal seismology
since it gives an acceptable error of around $10\%$
in, e.g., probing the coronal magnetic field based on Equation \eqref{eq_WKB}.

\subsection{Resonant damping and lateral leakage}
\label{subsec_leakage}

   \begin{figure}
   \centering
   \includegraphics[width=1.0\hsize]{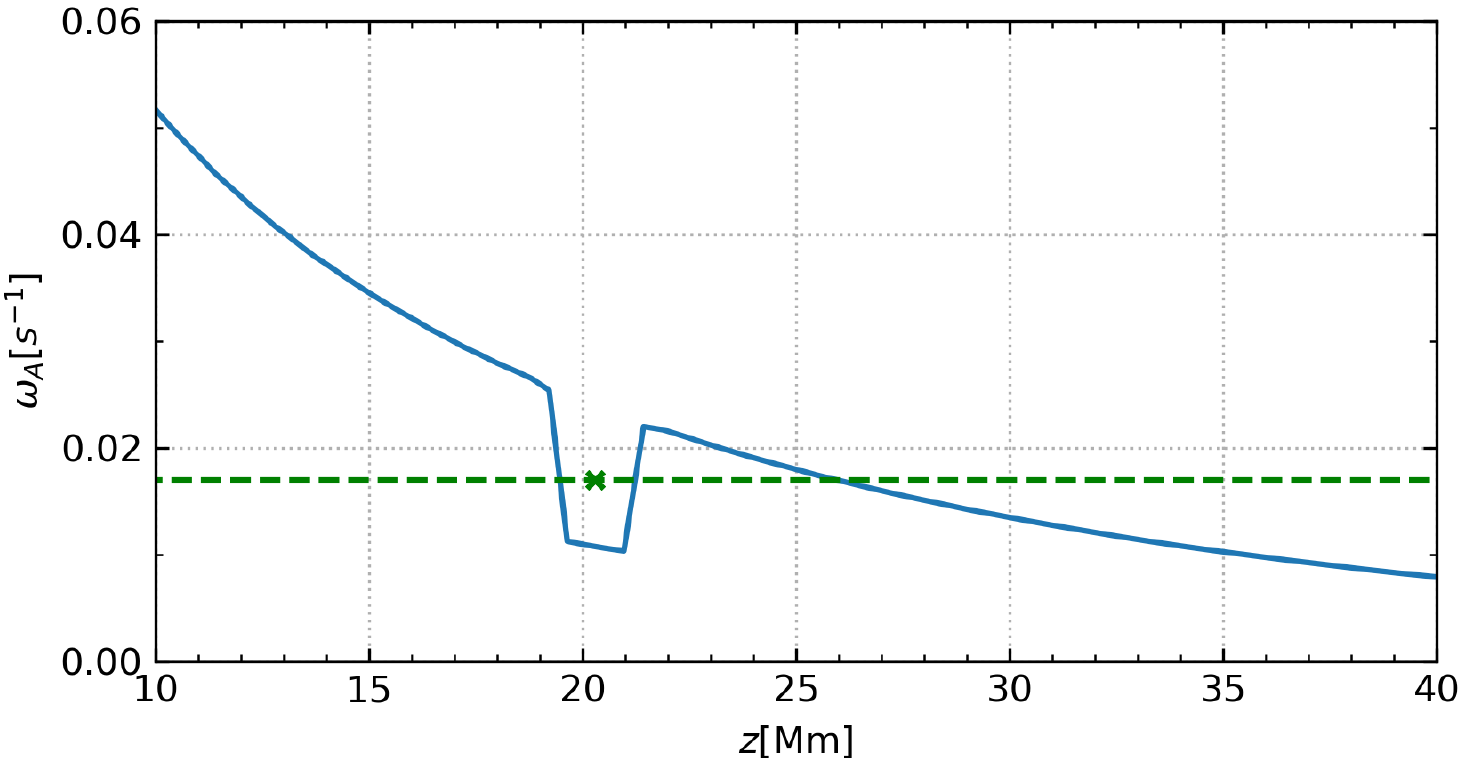}
   \caption{Alfv\'en frequency variation along the $z$-direction at $x=0, y=0$. The cross maker represents the predicted eigenfrequency of the kink mode in the curved loop, which is calculated from Equation~\eqref{eq_WKB}.}
    \label{fig_alfven_freq}
    \end{figure}  

   \begin{figure*}
   \centering
   \includegraphics[width=1.0\hsize]{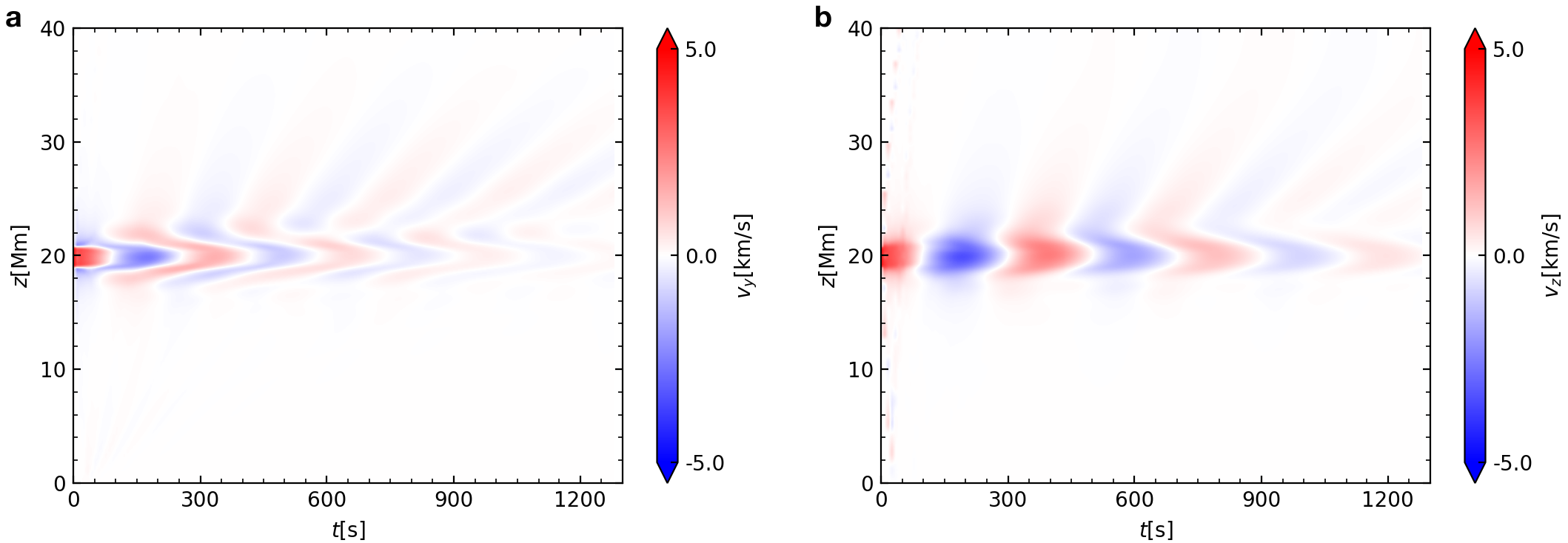}
   \caption{The time-distance map of velocity for (a) horizontal and (b) vertical polarization. The evolution of $v_y$ and $v_z$ over time is sampled at $x=0$, $y=0$ along the $z$-direction.}
    \label{fig_vel_td}
    \end{figure*}  
    
Resonant absorption can be observed in the current oscillating loop.
It is well-known that the oscillation amplitude decreases due to the
resonant energy transfer from the collective kink modes to the local 
Alfv\'en waves \citep[e.g.,][]{2020ApJ...904..116G}.
In Figure~\ref{fig_vyz_t},
we can indeed observe the damping of oscillations in the loop region.
Meanwhile,
wave leakage can also be captured,
as shown in the bottom panels of Figure~\ref{fig_vh} and Figure~\ref{fig_vv}.
Thus a comparison between the damping efficiency of resonant absorption
and lateral leakage is necessary.

Before proceeding,
we first revisit the physics of lateral leakage in a curved loop.
As discussed in \citet{2006A&A...446.1139V},
under a slab model,
the oscillation property depends on the slope of the Alfv\'en 
frequency profile.
The local Alfv\'en frequency can change with height,
due to the variation of magnetic field strength or vertical stratification 
of density. 
Let us consider the upward leakage.
Once the local Alfv\'en frequency decreases with height and becomes smaller than the wave frequency at a certain region above an oscillating loop,
the wave energy leaks out from the loop region through tunnelling the evanescent barrier,
leading to an oscillatory solution at a larger distance above the oscillating loop.
In our current model,
the local Alfv\'en frequency along the $z$-direction is of the form
\begin{eqnarray}
\omega_{\rm A}(z) = 
k(z) v_{\rm Ai}(z)=\displaystyle\frac{\pi}{L(z)}\displaystyle\frac{B(z)}{\sqrt{\mu_0 \rho(z)}},
\label{eq_alfven_freq}
\end{eqnarray}
where the local wavenumber $k(z)$ changes 
due to the variation of field line length $L(z)$.
Here,
$B(z)=B_0\exp{(-z/\Lambda_B)}$ represents the magnetic field strength
at a given height $z$. 
Given that the magnetic field strength also drops with $z$,
so ${\rm d} \omega_{\rm A}/{\rm d}z < 0$.
Therefore,
we should expect an evanescent barrier and the wave tunnelling effect. 
Figure~\ref{fig_alfven_freq} shows the variation of Alfv\'en frequency 
$\omega_{\rm A}$ along the $z$-direction at the loop apex.
The green dashed line labels the predicted oscillating frequency of the loop according
to Equation \eqref{eq_WKB}.
Indeed, it reveals an evanescent barrier around $21{\rm Mm}\lesssim z\lesssim26{\rm Mm}$
and smaller Alfv\'en frequencies than the eigenfrequency when $z\gtrsim 26$Mm.
The location of the evanescent barrier in the numerical models may slightly differ from this prediction,
due to the deviation of the eigenfrequencies of kink polarizations from the analytical value.
    
The oscillatory patterns shown in the upper loop region are now understandable with the aid of Figure~\ref{fig_alfven_freq}.
For clear illustration,
we also consider a time-height map of velocity at $y=0$ of the loop apex,
as shown in Figure~\ref{fig_vel_td}.
Oblique stripes above $z\sim23{\rm Mm}$ can be observed in both polarizations.
In Figure~\ref{fig_alfven_freq},
the Alfv\'en frequency decreases and becomes smaller than the predicted eigenfrequency when $z\gtrsim 26$Mm (it should be $z\gtrsim 23$Mm in the horizontal case due to the eigenfrequency deviation),
leading to a larger period with height in the upper loop region.
Therefore,
we could see the oblique stripes become less vertical with time.
Similar cross-field wave propagation can also be observed in e.g., \citet[][]{2015ApJ...812..121K,2017A&A...602A..75R} due to phase mixing.
For the horizontal polarization,
the evanescent barrier where the oscillations should be evanescent is mixed with the velocity signals of resonant Alfv\'en waves.
Nonetheless,
we can still observe a non-velocity region around $z\sim23{\rm Mm}$ 
after about $t=500$s.
In the case of vertical polarization,
however,
the evanescent patterns are not visible until $t\gtrsim900$s.
The evanescent barrier is probably covered by
the extension of the external velocity field, say, the wing of kink oscillations 
in the $z$-direction.

\begin{figure}
   \centering
   \includegraphics[width=1.0\hsize]{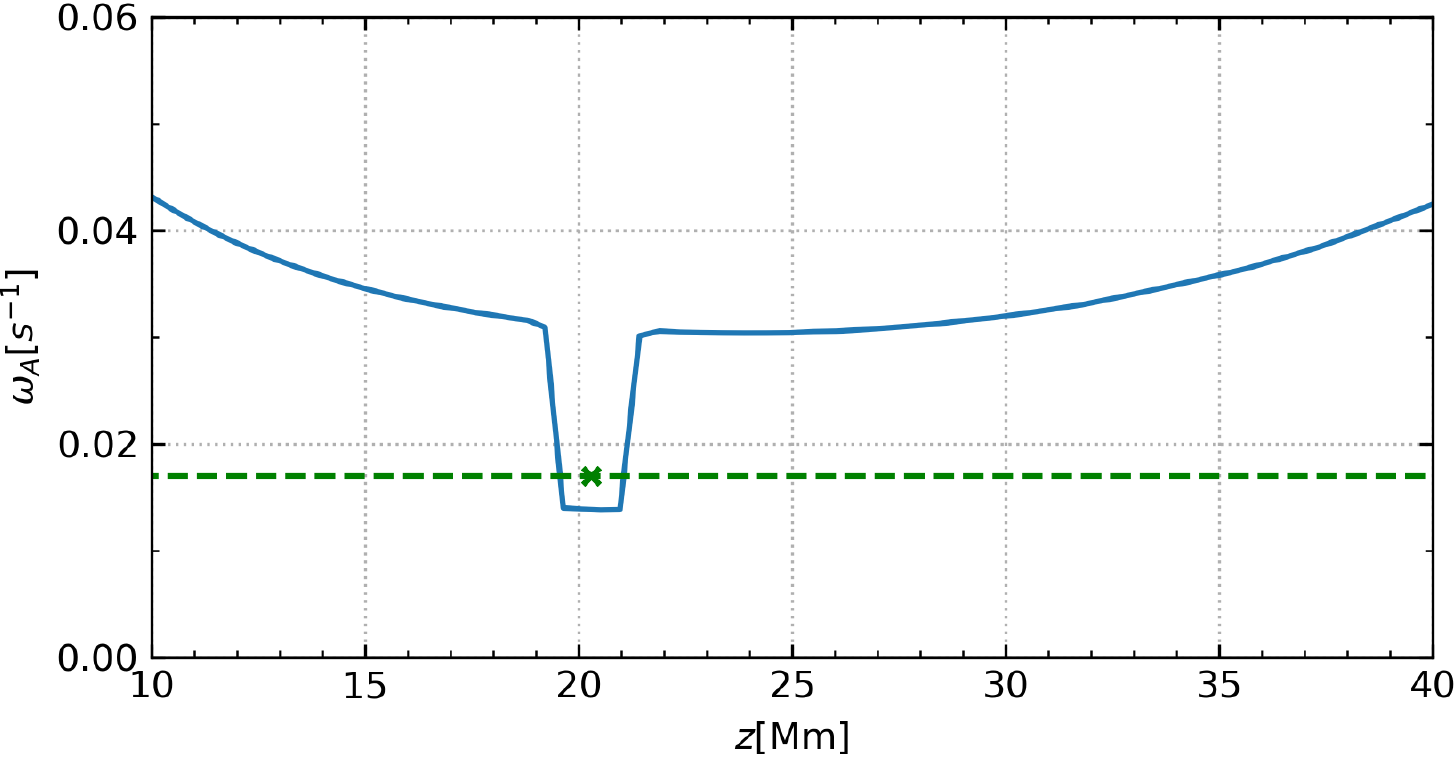}
   \caption{Similar to Figure~\ref{fig_alfven_freq}, but for the density stratified loop. The cross marker represents the predicted eigenfrequency of the kink mode according to Equation \eqref{eq_WKB}.}
    \label{fig_alfven_freq2}
    \end{figure}

\begin{figure*}
   \centering
   \includegraphics[width=1.0\hsize]{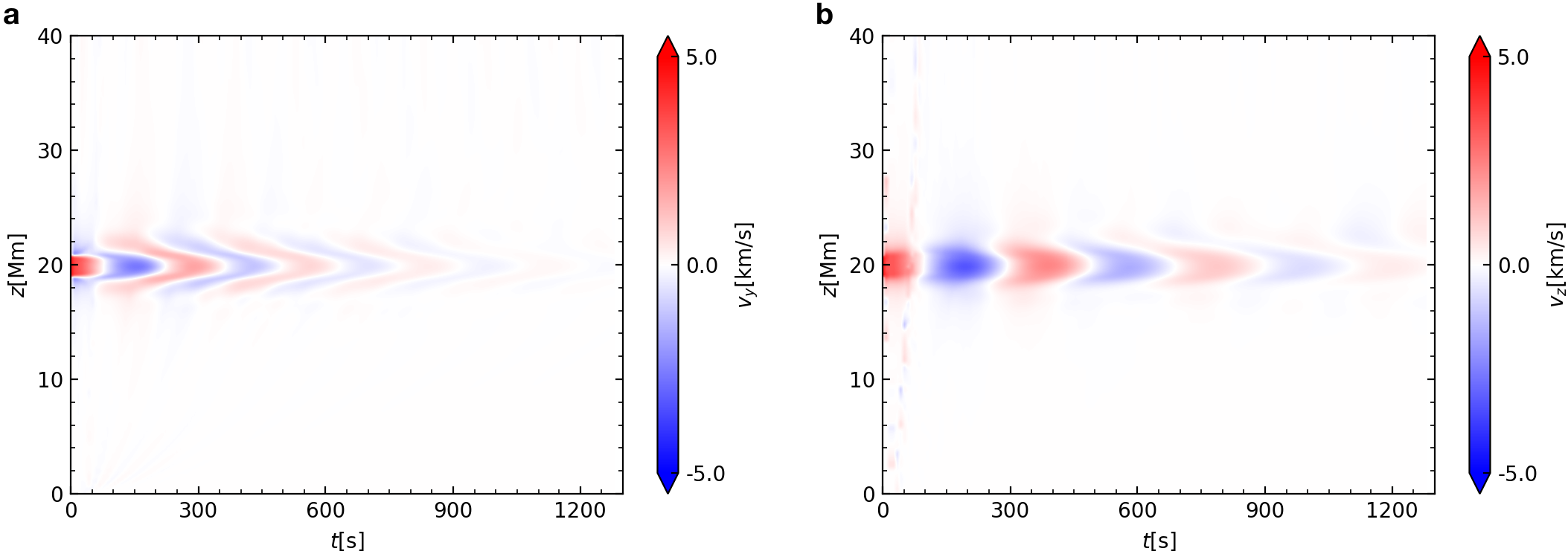}
   \caption{Similar to Figure~\ref{fig_vel_td}, but for the density stratified loop.}
    \label{fig_vel_td2}
    \end{figure*}

To quantitatively examine the effect of lateral leakage in the current model,
a straightforward idea is to isolate the resonant absorption by considering a non-leakage model.
In line of this,
a vertically stratified density distribution is considered.
The Equation~\eqref{eq_rho} is then replaced by
\begin{eqnarray}
\rho'(\Bar{r}) = \rho'_
{\rm e}+(\rho'_{\rm i}-\rho'_{\rm e})f(\Bar{r}),
\label{eq_rho_prime}
\end{eqnarray}
where
\begin{eqnarray}
\rho'_{\rm i}=2\rho_{\rm i}\exp\left[{-\displaystyle\frac{(z/L_0)^{\mu}}{\Lambda_B/L_0}}\right],~
\rho'_{\rm e}=2\rho_{\rm e}\exp\left[{-\displaystyle\frac{(z/L_0)^{\mu}}{\Lambda_B/L_0}}\right]
\label{eq_rhoe_prime}
\end{eqnarray}
with $L_0=10$Mm being the unit of length,
and the index $\mu=1.8$.
The density structure of this stratified loop is shown in Figure~\ref{fig_loop_snapshot}b.
For a more reasonable comparison,
this stratified model ensures the same eigenfrequency as the 
uniform density model.
The eigenperiod can be estimated using the WKB approximation
given by Equation~\eqref{eq_WKB},
but with the internal Alfv\'en speed defined by 
\begin{eqnarray}
v_{\rm Ai}(s)=\displaystyle\frac{B(s)}{\sqrt{\mu_0 \rho(s)}},
\label{eq_WKB_alfven2}
\end{eqnarray}
where $\rho(s)$ is the density variation along $s$.
In this model,
the local Alfv\'en frequency along the $z$-direction is illustrated
in Figure~\ref{fig_alfven_freq2}.
We can see the Alfv\'en frequency outside the loop is always larger
than the eigenfrequency (dashed line),
resulting in the absence of the evanescent barrier in this model.
In an intuitive manner,
we can compare the time-distance map of velocity for both polarizations with Figure~\ref{fig_vel_td},
as shown in Figure~\ref{fig_vel_td2}.
Not surprisingly, 
the patterns in the upper external loop region in Figure~\ref{fig_vel_td} disappear in Figure~\ref{fig_vel_td2} for both polarizations,
indicating that no wave leakage occurs in this stratified loop.
We can thus attribute any damping in this model to resonant absorption.  

Note that we only examine the local Alfv\'en frequency profile at the loop apex ($x=0$).
In Equation \eqref{eq_WKB_alfven2},
we find the term $B(s)/\sqrt{\rho(s)}$ monotonically decreases with height lower than the loop apex.
This implies that the Alfv\'en frequency at other heights in the loop region is always larger than the eigenfrequency shown in Figure~\ref{fig_alfven_freq2}.
Thus the evanescent barrier is indeed absent in the entire loop region.

   \begin{figure}
   \centering
   \includegraphics[width=1.0\hsize]{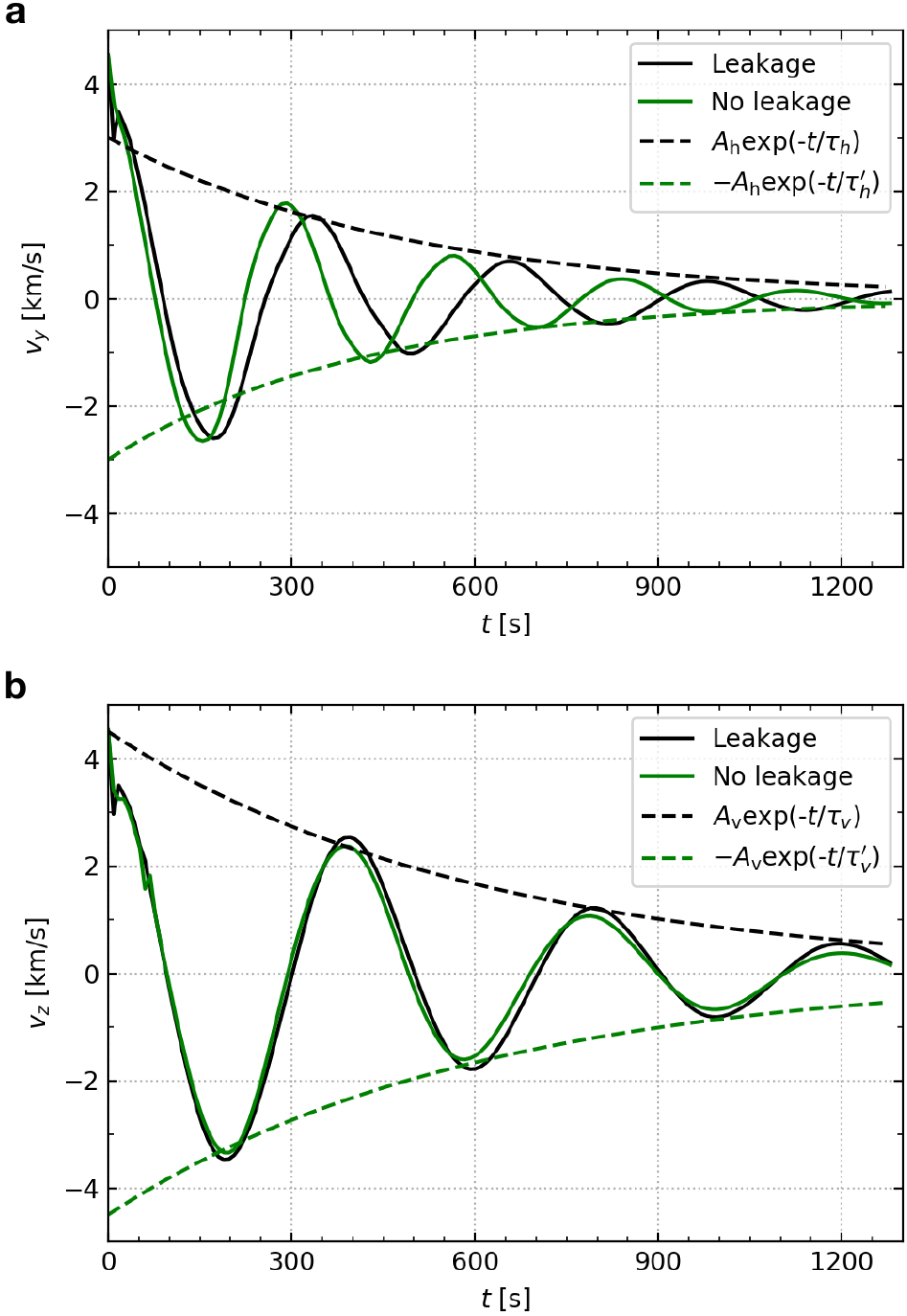}
   \caption{The oscillating profiles of the (a) horizontally and (b) vertically polarized modes in all models. The velocity is sampled at loop apex ($x=0, y=0, z=19.85{\rm Mm}$). Oscillations with and without wave leakage are distinguished by different colours. Dashed lines represent fitting results with $A_{\rm h}=3{\rm km/s}$ and $A_{\rm v}=4.5{\rm km/s}$ in the exponential functions.}
    \label{fig_vel_fit}
    \end{figure} 

Figure~\ref{fig_vel_fit} displays the fitting results for oscillating profiles in all models. 
Using the similar procedure described in Section~\ref{subsec_frequency},
we obtain a period of $P^{\prime}_{\rm h}=274.8$s for the horizontal polarization in the stratified loop.
This period shows a deviation of about $25.8\%$, 
compared with the predicted eigenfrequency $P_{\rm k}$.
This discrepancy will be discussed in detail in section ~\ref{subsec_freq_diff}.
An exponential function of the form $A\exp{(-t/\tau)}$ is employed for the fit,
represented by black (leakage model) and green (non-leakage model) dashed lines in Figure~\ref{fig_vel_fit}.
The fitting procedure yields a damping-time-to-period ratio $\tau_{\rm h}/P_{\rm h}=1.5$
($\tau_{\rm v}/P_{\rm v}=1.5$)
for the oscillating loop with lateral leakage and 
$\tau^{\prime}_{\rm h}/P^{\prime}_{\rm h}=1.5$
($\tau^{\prime}_{\rm v}/P^{\prime}_{\rm v}=1.5$)
for the loop without wave leakage.
For comparison,
we use the expression for the damping time due to resonant absorption under the thin-tube-thin-boundary (TTTB) approximation.
For a transversely linear density distribution \citep[e.g.,][]{2009A&A...503..213G,2013ApJ...777..158S},
it gives,
\begin{eqnarray}
\tau=\displaystyle\frac{2d}{\pi^2 l}\displaystyle\frac{\rho_{\rm ie}+1}{\rho_{\rm ie}-1}P_{\rm k},
\label{eq_TTTB_tau}
\end{eqnarray}
where $d$ represents the width of the loop.
If we consider the loop width varying from $d=a$ at the footpoint to $d=2a$ at the loop apex,
we can readily estimate the damping-time-to-period ratio to be $\tau/P_{\rm k}=1.01-2.03$.
This implies that our fitted damping time is very close to the analytically derived value for the damping time due to resonant absorption.
Note that this is a rough estimation,
as the damping rate may vary with different transverse density profiles,
as discussed in many previous studies \citep[e.g.,][]{2002ApJ...577..475R,2009A&A...503..213G}.
Nonetheless,
the same damping-time-to-period ratio in both uniform-density and stratified loop models 
indicates that the damping effect due to lateral leakage is not significant.
This probably means that the damping time of lateral leakage is much longer than that of resonant absorption.
This quantitatively confirms that lateral leakage is less efficient as a damping mechanism in curved coronal loops,
compared with resonant absorption.

In addition,
the same damping-time-to-period ratio for both polarizations indicates that the damping caused by resonant absorption does not have a preferred direction. 
Thus the significant difference in the number of observational events of horizontal and vertical kink polarizations may just be due to the preference of possible external exciters (coronal eruptions/ejections as suggested by \citealt{2015A&A...577A...4Z}). 
In other words,
the exciters of kink modes primarily interact with loops in horizontal directions,
rather than vertical directions.
Additionally,
the same damping-time-to-period ratio in the stratified model indicates that the rare occurrence of vertical polarizations should not be attributed to the density stratification.

Note that our current analysis only considers the exponential damping stage.
Previous studies \citep[e.g.,][]{2012A&A...539A..37P,2013A&A...551A..39H,2016A&A...589A.136P,2016A&A...595A..81M,2020ApJ...904..116G}
have confirmed that the damping profile of kink modes consists of two stages:
the Gaussian stage and the subsequent exponential stage.
We can see that the first period of the oscillation profiles in Figure~\ref{fig_vel_fit} can not be perfectly described by exponential functions,
probably suggesting a Gaussian damping stage,
which is known to robustly manifest in non-axisymmetric loops \citep{2020ApJ...904..116G}.
In the current analysis, 
however,
we only focus on the exponential damping profile for a more direct comparison with the thin-tube-thin-boundary (TTTB) expression,
as discussed in \citet{2020ApJ...904..116G}

   \begin{figure}
   \centering
   \includegraphics[width=1.0\hsize]{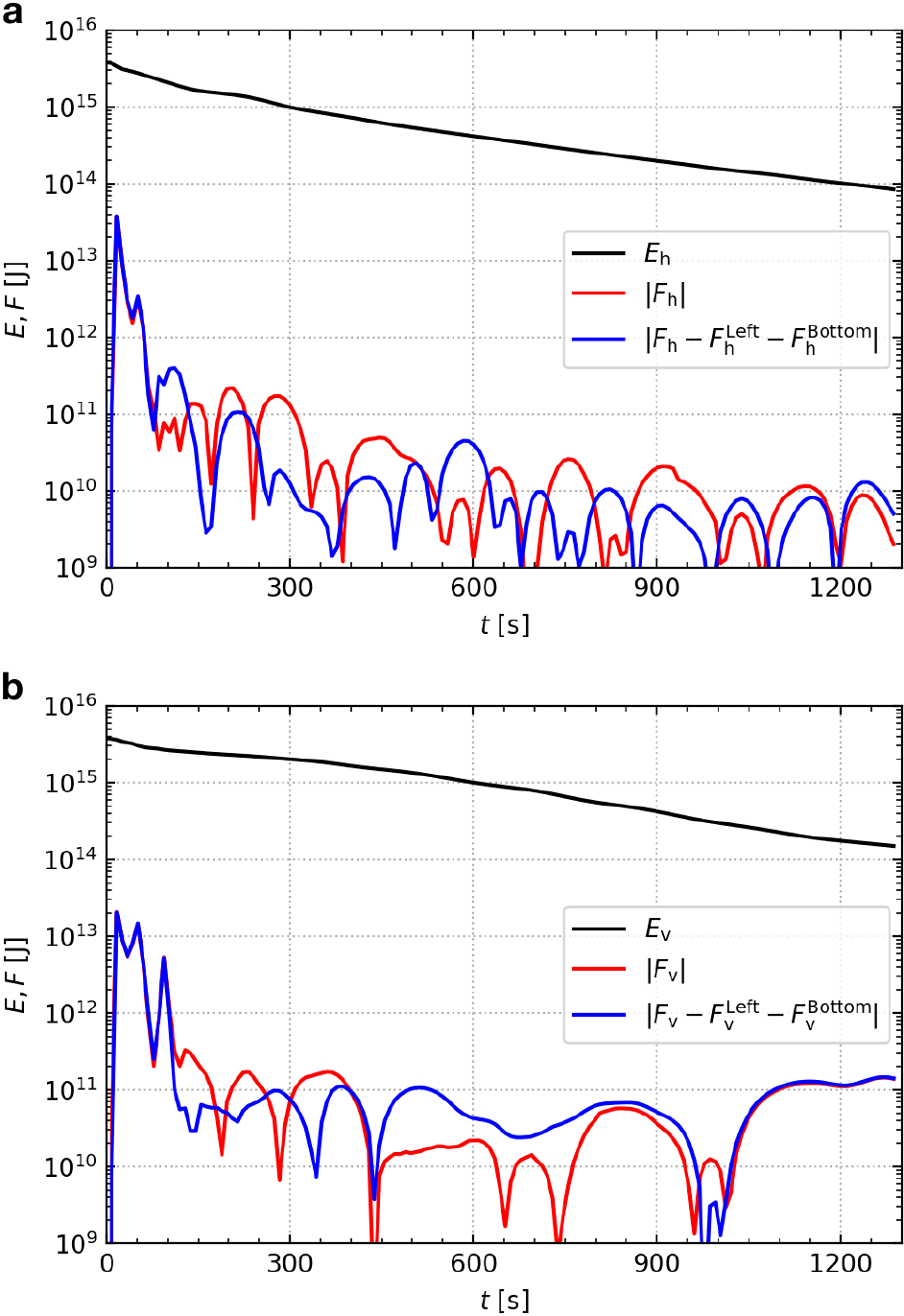}
   \caption{Energy and energy flux densities given by Equation~\eqref{eq_E} and Equation~\eqref{eq_F} for (a) horizontal and (b) vertical polarizations.
   Here the subscript "h" ("v") represents the horizontal (vertical) case.
   The energy densities for both polarizations are depicted with black lines.
   Energy flux densities are computed in two ways: 
   by considering all boundaries and by excluding the left symmetric boundary and bottom reflective boundary.
   These two cases are distinguished by red and blue curves.}
    \label{fig_EF}
    \end{figure}  
    
Now we further compare the efficiency of resonant absorption and lateral leakage from the energetics perspective.
As shown above,
we would expect a scenario in which the damping time of
lateral leakage is much longer than that of resonant absorption.
In practice,
these time scales can be obtained by comparing 
the energy and energy flux densities in the computational domain.
As discussed in e.g., \citet[][]{2020ApJ...904..116G},
we can define wave-related energy and energy flux densities for linear perturbations
\begin{eqnarray}
E=\int_V \left(\displaystyle\frac{1}{2}\rho_0 \vec{v}_1^2+\displaystyle\frac{\vec{B}_1^2}{2\mu_0}+\displaystyle\frac{p_1^2}{2\gamma p_0}\right){\rm d}V,
\label{eq_E}
\end{eqnarray}
\begin{eqnarray}
F=\oint_{S} \left[p_1 \vec{v}_1
+ \frac{\vec{B}_1\times \left(\vec{v}_1\times\vec{B_0}\right)}{\mu_0}\right]\cdot{\rm d}\vec{S}.
\label{eq_F}
\end{eqnarray}
Here the subscript "1" represents the linear perturbation to the equilibrium.
$V$ represents the whole computational domain and 
$S$ is the surface of it.
Note that the source term defined by Equation~(B8) in \citet[][]{2020ApJ...904..116G} is not relevant in the current potential field given an initially constant pressure throughout.
In Figure~\ref{fig_EF},
we illustrate the evolution of total energy and energy flux densities in terms of the second-order small quantities given by Equation~\eqref{eq_E} and Equation~\eqref{eq_F}.
For a straightforward comparison,
we consider the amplitude of energy flux density variations, as shown by red and blue curves in Figure~\ref{fig_EF}.
Particularly,
the energy fluxes through open boundaries except the left symmetric and the bottom reflective boundaries should induce energy losses associated with lateral leakage.
In Figure~\ref{fig_EF},
the acceptable small difference between the red and blue curves reflects the efficiency of the symmetric and reflective boundaries in practice.
We find for both polarizations,
the total energy $E$ is much larger than the amplitude of the energy flux $|F|$,
and the time scale $E/|F|$ is larger than $10^3$s.
This means that the damping time associated with lateral leakage is about one order of magnitude longer than that of resonant absorption.
Therefore,
lateral leakage is less efficient as a damping mechanism, compared with resonant absorption.

The decrease in total energy in Figure~\ref{fig_EF} is probably numerical.
We recomputed the energy $E$ shown in Figure~\ref{fig_EF} by considering a lower AMR level (lower numerical resolution) and found a faster decrease with time. 
This hints that 
a larger grid spacing leads to an underestimation of the energy in the computational domain, 
particularly in the nonuniform boundary layer as discussed in \cite{2024arXiv240112885S}.
Nonetheless,
the statement that the damping time of lateral leakage is much larger than that of resonant absorption remains unchanged.

\subsection{Frequency discrepancies}
\label{subsec_freq_diff}

   \begin{figure}
   \centering
   \includegraphics[width=1.0\hsize]{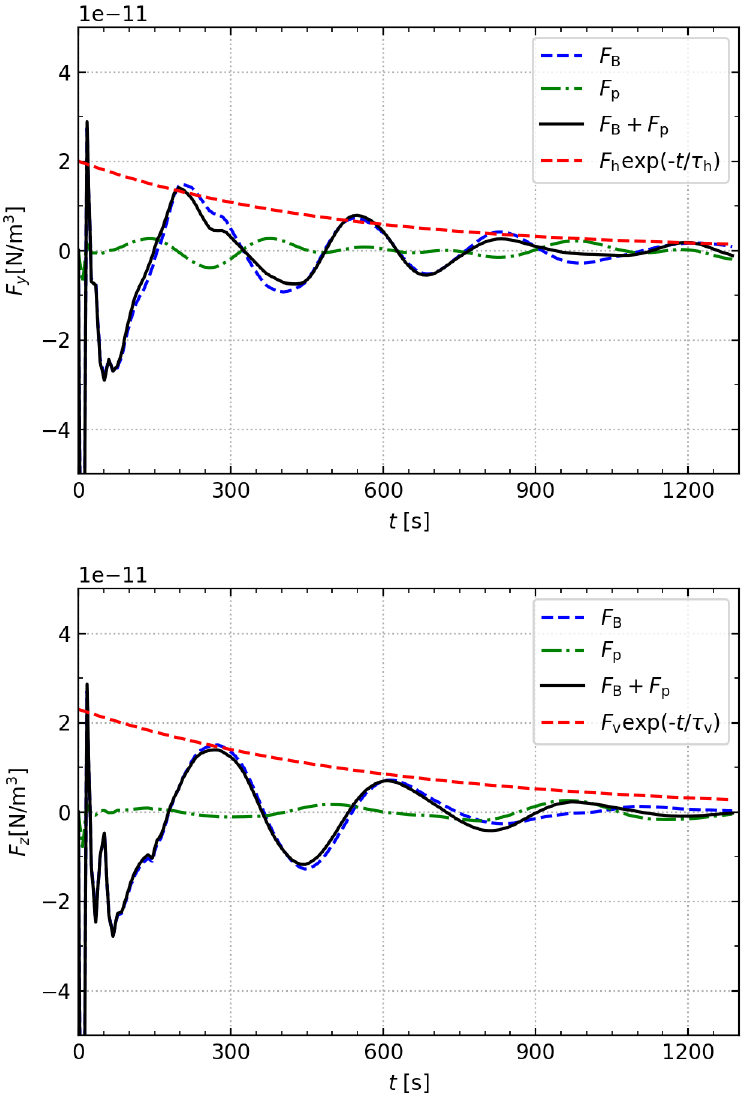}
   \caption{The evolution profiles of the magnetic force density (blue), the pressure gradient force density (green), and the resultant force density of these two (black). The corresponding forces are calculated by Equation~\eqref{eq_linearized_mom1} and Equation~\eqref{eq_linearized_mom2}.
   Dashed red lines represent fitting results with $F_{\rm h}=2.0 \times 10^{-11}{\rm N/m^3}$ and $F_{\rm v}=2.3 \times 10^{-11}{\rm N/m^3}$ in the exponential functions.}
    \label{fig_force1}
    \end{figure} 

   \begin{figure}
   \centering
   \includegraphics[width=1.0\hsize]{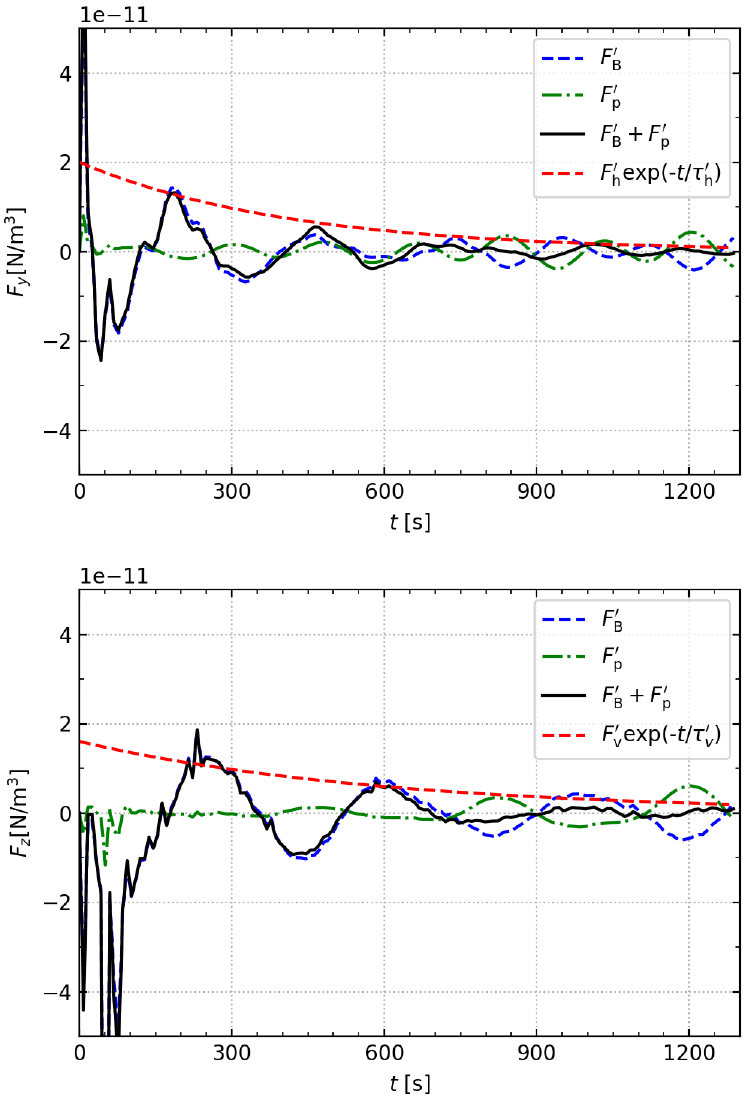}
   \caption{Similar to Figure~\ref{fig_force1}, but for the density stratified model. 
   Dashed red lines represent fitting results with $F^{\prime}_{\rm h}=2.0 \times 10^{-11}{\rm N/m^3}$ and $F^{\prime}_{\rm v}=1.6 \times 10^{-11}{\rm N/m^3}$ in the exponential functions.}
    \label{fig_force2}
    \end{figure} 

Now we attempt to understand period discrepancies shown up in different models.    
Seen from Figure~\ref{fig_vyz_t},
we can find a frequency discrepancy between the two polarizations.
Likewise,
the oscillation periods for the uniform density model and the stratified model are different as well,
as shown in Figure~\ref{fig_vel_fit}a.
These differences can be understood intuitively.
For instance,
the internal density decreases in the $z$-direction in the stratified loop,
leading to different dynamics across the loop in the $\hat{\vec{e}}_v$ direction.
While the uniform density loop has the same density distribution across the loop in the same direction.
Therefore,
the period difference between the uniform density model and the stratified model is reasonable since the density distributions perpendicular to the oscillating direction are different.
Following this line of thought,
we would expect no difference 
between the two periods of vertical polarization,
given that the density has no variation in the $y$-direction
across the loop.
This means that the vertically oscillating loop can not "feel" any difference in the environment from the left to the right in the $y$-direction. 
Indeed,
Figure~\ref{fig_vel_fit}b shows that the period of the vertical polarization is $P^{\prime}_{\rm v}=401.8$s,
which is the same as the period of vertical polarization in the uniform density loop model. 
This hints that the WKB approximation can only reasonably describe the eigenfrequency of kink modes in a loop without an asymmetric cross-loop density profile perpendicular to the oscillation direction.
The asymmetric cross-loop density profile can result in different dynamics from one side of the cross-section to the other,
leading to a deviation of the eigenfrequency from analytical predictions.

To further quantitatively understand the period discrepancies,
we consider the force analysis by employing the linearized momentum equation,
\begin{eqnarray}
\rho_0\displaystyle\frac{\partial \vec{v}}{\partial t}=F_{\rm p}+F_{\rm B},
\label{eq_linearized_mom1}
\end{eqnarray}
where
\begin{eqnarray}
 F_{\rm p}=-\nabla p_1,
 F_{\rm B}=\displaystyle\frac{\left(\nabla\times\vec{B}_1\right)\times\vec{B}_0}{\mu_0}+\displaystyle\frac{\left(\nabla\times\vec{B}_0\right)\times\vec{B}_1}{\mu_0}.
\label{eq_linearized_mom2}
\end{eqnarray}
Here $p_1$ represents the linear perturbation to the equilibrium pressure,
and $\rho_0=\rho(\Bar{r})$.
We calculate the average force density in a large circular region ($\sqrt{y^2+(z-19.85)^2}<5$Mm) at the apex ($x=0$).
Figure~\ref{fig_force1} and Figure~\ref{fig_force2} illustrate the profiles of the pressure gradient force density $F_{\rm p}$ (blue), the magnetic force density $F_{\rm B}$ (green), and the sum of these two.
We observe that for all cases,
especially for the leakage model (uniform density loop),
the resultant force is very close to the magnetic force until the amplitude of the forces becomes too small to recognize.
This means that the magnetic force \footnote{In fact, the magnetic tension component is dominant in our case,
given that the oscillation frequency is close to the local Alfv\'en frequency. As shown by Equation (5) in \citet[][]{2009A&A...503..213G},
the magnetic pressure component can be neglected when the oscillation frequency approaches the local Alfv\'en frequency.} is still the dominant restoring force. 
This agrees with what we already know about kink modes. 
Let us now focus on the resultant forces that are practically relevant in the current case.
By fitting all the resultant force profiles (black curves) in Figure~\ref{fig_force1} and Figure~\ref{fig_force2},
we can obtain the corresponding amplitudes.
In the uniform density loop,
the amplitude of the resultant force for the horizontal polarization is
$F_{\rm h}=2.0 \times 10^{-11}{\rm N/m^3}$,
while the corresponding amplitude for the vertical polarization is
$F_{\rm v}=2.3 \times 10^{-11}{\rm N/m^3}$.
Considering the amplitudes of the velocity profiles are $A_{\rm h}=3{\rm km/s}$ and $A_{\rm v}=4.5{\rm km/s}$, respectively,
we can readily deduce the frequency difference $\omega_{\rm h}/\omega_{\rm v}=1.29$, 
according to a Fourier decomposition in time of Equation~\eqref{eq_linearized_mom1}.
Therefore, 
the period difference between the horizontal and vertical polarizations seems understandable.
Likewise,
we obtain the corresponding amplitude of the fitting for the horizontal polarization is
$F^{\prime}_{\rm h}=2.0 \times 10^{-11}{\rm N/m^3}$ in the density stratified loop.
But be aware that the average density at the loop apex surface decreases about $25\%$.
Therefore,
the period deviation of about $25\%$ can be observed for the horizontal polarization in the stratified loops according to Equation~\eqref{eq_linearized_mom1}.
In addition,
the amplitude of the resultant force for the vertical polarization is $F^{\prime}_{\rm v}=1.6 \times 10^{-11}{\rm N/m^3}$
for the stratified loop,
thus the periods show almost no difference for both uniform density and stratified models.

\section{Discussion and Conclusions}
\label{sec_summary}

We investigated the horizontally and vertically polarized kink eigenmodes excited by initial velocity perturbations in curved coronal loops embedded in a potential magnetic field.
In the case of a uniform-density loop,
we found that the eigenfrequencies of both polarizations deviate from the WKB approximation by about $10\%$.
This implies that the WKB approximation effectively describes both horizontal and vertical polarizations of the kink eigenmodes in this uniform-density loop.
Lateral leakage of both polarizations was clearly observed in the upper loop region as expected.
For comparison,
we also considered a density-stratified loop
without an evanescent barrier,
resulting in the absence of wave leakage.
The damping-time-to-period ratios of both polarizations in the density-stratified loop were similar to those in the uniform-density loop,
indicating that wave leakage does not significantly influence the current damping of both polarizations.
All damping rates closely aligned with the predictions of the thin-tube-thin-boundary (TTTB) approximation,
indicating that resonant absorption is the dominant damping mechanism in the current curved loops.
Examining the oscillating frequency of both polarizations in the stratified loop,
we found that the eigenfrequency of the vertical polarization can be described by the WKB approximation,
while the frequency of the
horizontal polarization deviates from the WKB approximation by around $25\%$.
This implies that the WKB approximation is effective only in describing kink modes in a loop without
an asymmetric cross-loop density variation in the direction perpendicular to the polarization,
the vertical polarization for instance.

The deviations from the WKB approximation in the frequencies are reasonable.
\citet{2020ApJ...894L..23M} have pointed out that
the WKB approximation can well describe the eigenfrequency of horizontal kink oscillations
when taking into account the expansion of the loop due to the chosen potential magnetic field and the density stratification.
In our current model,
however,
a deviation of about $11\%$ was found in the horizontal polarization,
even if the changes in the loop cross-section are included,
as seen in Equation~\eqref{eq_TTTB_freq2} and Equation~\eqref{eq_WKB}.
Such a deviation may be caused by several reasons.
One is that our current loop is not long enough, thus leading to a deviation from the long wavelength limit ($kR\ll1$ with $k$ being the axial wave number and $R$ being the loop radius) considered in deriving Equation \eqref{eq_TTTB_freq}.
In addition,
we assume a uniform magnetic field across the loop (i.e., pressureless condition) when performing the WKB approximation.
Yet
the deviation is only around $10\%$.
This suggests that estimating parameters via seismology may not require measuring the external magnetic field.
However, it is worth noting that
the measurement of the internal Alfv\'en speed is still necessary for seismology practices, 
as indicated in Equation~\eqref{eq_WKB}.
Directly using Equation~\eqref{eq_TTTB_freq2} based on the thin-tube (TB) theory by assuming an average Alfv\'en frequency profile is still possible to induce significant errors.

 The frequency discrepancy between the two polarizations in the uniform loop model seems to contradict previous analytical and numerical findings \citep{2003A&A...409..287R, 2020ApJ...904..116G}.
In previous studies,
a scenario has been discussed in which
a larger (smaller) axis in the loop cross-section results in a smaller (larger) frequency of kink modes,
given that more contribution of the external (internal) Alfv\'en frequency is involved in the minor (major) axis. 
In the present model, 
the radius of the loop cross-section in the horizontal direction
is fixed,
while the radius of the loop cross-section in the vertical direction increases from the bottom to the loop apex.
However,
the frequency of the horizontal polarization
is larger than that of the vertical polarization.
This contradiction is probably induced by the changing magnetic field strength with height in the present model.
In this case,
the Alfv\'en frequency (or Alfv\'en speed if we neglect the axial wave number variation since the loop is thin) decreases from the lower loop region to the upper loop region,
which is the main difference from previous studies \citep[e.g.,][]{2003A&A...409..287R,2020ApJ...904..116G}.
In the current model,
we would expect the variation in Alfv\'en speed in the cross-section to influence the dynamics of the loop.
At a given height,
the Alfv\'en speed changes along the minor axis of the cross-section (i.e., the $\hat{e}_v$ direction).
For the vertical polarization,
such Alfv\'en speed variation in the oscillation direction induces the contraction or extension of the cross-section in the minor direction.
In the horizontally polarized case,
the variation in Alfv\'en speed is perpendicular to the oscillation direction, 
thus inducing an oblique minor axis.
In addition,
the variation in the Alfv\'en speed along the loop also affects the eigenfunctions of kink modes. 
Therefore,
the dynamics of the present loop are different from canonical elliptical loops.
It is not straightforward to compare directly with the scenario discussed in 
\cite{2003A&A...409..287R} and \cite{2020ApJ...904..116G}.

The density-stratified loop is artificially designed for comparison with the uniform-density loop.
To isolate the effect of resonant absorption,
the density-stratified loop is modified by adjusting the scale height determined by the index $\mu$,
as defined in Equation~\eqref {eq_rhoe_prime}.
This modification ensures the absence of an evanescent barrier above the loop due to a larger Alfv\'en frequency,
compared with the uniform-density loop.
Meanwhile,
the density-stratified loop maintains the same predicted eigenfrequency as given by the WKB approximation.
This makes the comparison between the two types of oscillating loops more reasonable.
Even though lateral leakage is absent in the density-stratified model,
the damping-time-to-period ratio remains the same as in the uniform density loop.
This clearly shows that lateral leakage is less effective than resonant absorption.

A nonlinear regime has not been examined in the current study.
Given that both resonant absorption and lateral leakage are linear processes,
we thus consider a linear regime with small amplitude perturbations to excite different polarizations,
ensuring amplitudes significantly smaller than the loop width.
In this manner,
the damping of kink polarizations is not influenced by nonlinearity discussed in e.g., \citet[][]{2016A&A...590L...5G,2016A&A...595A..81M,2021ApJ...910...58V},
thereby making the comparisons with linear analytical theories more straightforward.

\begin{acknowledgements}
     The authors acknowledge the funding from the European Research Council (ERC) under the European Unions Horizon 2020 research and innovation program (grant agreement No. 724326). TVD was also supported by the C1 grant TRACEspace of Internal Funds KU Leuven and a Senior Research Project (G088021N) of the FWO Vlaanderen.
     Furthermore, TVD received financial support from the Flemish Government under the long-term structural Methusalem funding program, project SOUL: Stellar evolution in full glory, grant METH/24/012 at KU Leuven. The research that led to these results was subsidised by the EU through the DynaSun project (number 101131534 of HORIZON-MSCA-2022-SE-01) and the Belgian Federal Science Policy Office through the contract B2/223/P1/CLOSE-UP.
     M.G. and B.L. acknowledge support from the National Natural Science Foundation of China (12203030, 12373055, 41974200).
     The authors thank Norbert Magyar and Konstantinos Karampelas for helpful discussions.
\end{acknowledgements}

%
\bibliographystyle{aa} 
\bibliography{ref} 
%

\end{document}